\documentclass[apj]{emulateapj}
\usepackage{mathptmx}
\usepackage{amsmath, amsthm, amssymb}

\newcommand  \acc      {\ifmmode {\rm km\,s}^{-2} \else km\,s$^{-2}$\fi}
\newcommand  \kms      {\ifmmode {\rm km\,s}^{-1} \else km\,s$^{-1}$\fi}

\newcommand  \cmii     {\hbox{cm$^{-2}$}}
\newcommand  \ergs     {\ifmmode {\rm erg\,s}^{-1} \else erg s$^{-1}$\fi}
\newcommand  \ergcms   {\ifmmode {\rm erg\,cm}^{-2}\,{\rm s}^{-1}
                        \else erg\,cm$^{-2}$\,s$^{-1}$\fi}
\newcommand  \ergcmsA  {\ifmmode{\rm erg\,cm}^{-2}\,{\rm s}^{-1}\,{\rm\AA}^{-1}
                        \else erg\,cm$^{-2}$\,s$^{-1}$\,\AA$^{-1}$\fi}
\newcommand  \ergcmsHz {\ifmmode{\rm erg\,cm}^{-2}\,{\rm s}^{-1}\,{\rm Hz}^{-1}
                        \else erg\,cm$^{-2}$\,s$^{-1}$\,Hz$^{-1}$\fi}
\newcommand  \phcms    {\ifmmode {\rm ph\,cm}^{-2}\,{\rm s}^{-1}
                        \else ph\,cm$^{-2}$\,s$^{-1}$\fi}
\newcommand  \phcmsA   {\ifmmode {\rm ph\,cm}^{-2}\,{\rm s}^{-1}\,{\rm\AA}^{-1}
                        \else ph\,cm$^{-2}$\,s$^{-1}$\,\AA$^{-1}$\fi}

\newcommand  \pgt      {PG\,1211+143}
\newcommand  \xmm      {{\it XMM-Newton}}

\journalinfo{The Astrophysical Journal, 
636: ?1--?7, 2006 January, astro-ph}
\slugcomment{Received 2005 May 5; accepted 2005 September 17}

\shorttitle{THE HIGH-VELOCITY OUTFLOW OF PG\,1211+143}

\shortauthors{KASPI \& BEHAR}

\begin{document}

\title{The High-Velocity Outflow of PG\,1211+143 May Not be That Fast}

\author{
Shai~Kaspi\altaffilmark{1,2}
and
Ehud Behar\altaffilmark{1}
}

\altaffiltext{1}{Physics Department, Technion, Haifa 32000, Israel;
shai@physics.technion.ac.il, behar@physics.technion.ac.il}
\altaffiltext{2}{School of Physics and Astronomy and the Wise
Observatory, Raymond and Beverly Sackler Faculty of Exact Sciences,
Tel-Aviv University, Tel-Aviv 69978, Israel}

\begin{abstract}
We analyze the X-ray spectrum of the quasar PG\,1211+143 observed with
the CCD and grating spectrometers on board \xmm .  Using an ion by
ion fitting model we find an outflow component of about 3000~\kms\
that includes absorption lines of K-shell and L-shell ions of the
astrophysically abundant elements.  We also identify and include in our
model broad (FWHM = 6000\,\kms ) emission lines from H-like ions of C,
N, O, and Ne, and He-like ions of O, Ne, and Mg. The outflow velocity
we find is an alternative interpretation of the data and is in contrast
with the ultra high velocity of $\sim$~24000~\kms\ reported previously
for this object. Nevertheless, we can not completely rule out the
presence of a high velocity component due to the poor signal-to-noise
ratio of the data.
\end{abstract}


\section{Introduction}

The phenomenon of outflowing mass from the cores of Active Galactic
Nuclei (AGNs) has been well established by now. Typical velocities
of a few hundreds to a few thousands \kms\ have been measured in numerous
objects (Crenshaw et al. 2003 and references therein). Recent studies
of the X-ray spectra of certain quasars have led to claims of much
higher outflow velocities reaching a significant fraction of the
speed of light.  In APM\,08279+5255, Chartas et al. (2002) claim
to find absorption lines of \ion{Fe}{25}\,He$\alpha$ outflowing at
speeds of $\sim0.2c$ and $\sim0.4c$. Hasinger et al.  (2002) using
different instruments prefer a more conservative interpretation for
APM\,08279+5255, by which the X-ray wind is much slower and consistent
with the well known, UV broad absorption line wind of that source,
outflowing at velocities of up to 12000 \kms. In PG\,1115+080,
Chartas et al.  (2003) find two X-ray absorption systems which they
attribute to \ion{Fe}{25}\,He$\alpha$ and hence to outflow velocities
of $\sim0.10c$ and $\sim0.34c$.

All of these measurements, however, were carried out using spectra
obtained with CCD cameras and hence at moderate spectral resolving
powers of $R\sim 50$. Using the simultaneous observing mode of \xmm,
which operates both CCD cameras and reflection gratings ($R$ up to 500)
simultaneously, high resolution X-ray spectra for several quasars have
been obtained.  For \pgt\ Pounds et~al. (2003a) find a rich, well
resolved spectrum featuring absorption lines of several ions, which
they interpret as due to an outflow of $\sim$~24000~\kms. A similar
interpretation was applied to a similar observation of PG\,0844+349,
where Pounds et al. (2003b) report even higher velocities reaching
$\sim$~60000~\kms. In NGC\,4051, Pounds et al. (2004) find a
single absorption line at $\sim$~7.1~keV, which they suggest may be
\ion{Fe}{26} Ly$\alpha$ at an outflow velocity of $\sim$~ 6500~\kms,
or the He$\alpha$ resonance absorption line of \ion{Fe}{25} in
which case the outflow velocity is $\sim$~16500~\kms. Yet another
ultra-high-velocity  (UHV, i.e., sub-$c$) wind of 50000~\kms\ was
reported by Reeves et al. (2003) for PDS~456. In all of these sources,
the inferred hydrogen column densities through the wind is of the
order of 10$^{23}$~\cmii, which is about an order of magnitude higher
than the typical values measured for the nearby Seyfert sources.

If indeed UHV outflows are common to bright quasars, this could have
far reaching implications on our understanding of AGN winds and AGNs
in general. For instance, if these winds carry a significant amount
of mass as the high column densities may suggest, they would alter
our estimates of the metal enrichment of the intergalactic medium by
quasars.  It remains to be shown theoretically what mechanism (e.g.,
radiation pressure) can drive these intense winds. Since the amount
of mass in the wind is not well constrained, it is still unclear what
effect it may have on the energy budget of the AGN.  King \& Pounds
(2003) note that UHV winds have been found mostly for AGNs accreting
near their Eddington limit. They provide a theory by which the UHV
outflows are optically thick producing an effective photosphere,
which is also responsible for the UV blackbody and soft X-ray (excess)
continuum emission observed for these sources.

Due to the potentially profound importance of the UHV wind phenomenon,
in the present work, we re-analyze the \xmm\ observation of \pgt\
first studied in Pounds et al. (2003a). Using a significantly
different method, we propose alternative identifications of the
absorption lines detected by Pounds et al.  The present analysis
suggests that the outflow velocity of the absorber in \pgt\ may
be more similar to the common outflows found in Seyfert galaxies.
In \S~2 and \S~3 we describe the data reduction and the data analysis,
respectively, and in \S~4 we discuss our model and results. In \S~5 we
give a short summary of our study. Throughout this paper we use for
\pgt\ a redshift of $z=0.0809$ (Marziani et al. 1996) and a Galactic
absorption column density of $2.85\times10^{20}$ \cmii\
(Murphy et al. 1996). The bolometric luminosity of \pgt\ is estimated
to be about $4.5\times 10^{45}$ \ergs\ (Kaspi et al. 2000).

\section{Data Reduction}
\label{datared}

\pgt\ was observed with {\it XMM-Newton} during 2001 June 15 for
about 55 ks. We retrieved the data for this observation from the
{\it XMM-Newton} archive and reduced them using the Science Analysis
System (SAS v5.3.0) in the standard processing chains as described
in the data analysis threads and the ABC Guide to {\it XMM-Newton}
Data Analysis.\footnote{http://heasarc.gsfc.nasa.gov/docs/xmm/abc}

The EPIC (CCD) data were processed using SAS in the standard way.
Source data were extracted from circular regions of radius
60$\arcsec$. The EPIC-pn was operated in the large window mode
resulting in a net exposure time of 50 ks. The observed count rate
($\sim$3.7 counts\,s$^{-1}$ before background subtraction) was well
below the pileup threshold of 12 counts\,s$^{-1}$ (see the \xmm\
Users' Handbook).  For statistical purposes, we binned the spectra
to have at least 25 counts per bin.

The RGS1 and RGS2 were operated in the standard spectroscopy mode
resulting in an exposure time of $\sim 52$ ks.  Background subtraction
was performed with the SAS using regions adjacent to those containing
the source in the spatial and spectral domains. The spectra were
extracted into uniform bins of $\sim$\,0.04 \AA\ (which is about
the RGS resolution and is 4 times the default bin width) in order
to increase the signal-to-noise ratio (S/N). For the purpose of
modeling narrow absorption lines, this rebinning method is better
than the method used by Pounds et al. (2003a) of rebinning the
spectrum to a minimum of 20 counts per bin, which distorts the
spectrum especially around low-count-rate absorption troughs. To
flux calibrate the RGS spectra we divided the count spectrum of
each instrument by its exposure time and its effective area at each
wavelength. Each flux-calibrated spectrum was corrected for Galactic
absorption and the two spectra were combined into an error-weighted
mean. At wavelengths where the RGS2 bins did not match exactly the
wavelength of the RGS1 bins, we interpolated the RGS2 data to enable
the averaging. We carefully checked that the features seen in the
combined spectrum appear in both RGS1 and RGS2. In several pixels
which show large deviation in counts due to incomplete rejection of
some warm pixels (J. Kaastra, private communication) we used the
other RGS.  The sky-subtracted combined RGS spectrum has in total
$\sim 8900$ counts and its S/N ranges from $\sim 2$ around 8\,\AA\
to $\sim 5$ around 18\,\AA\ with an average of 3.  Statistics in the
second order of refraction are insufficient, hence we did not include
it in our analysis.

Overall, the present data reduction agrees well with that of Pounds
et al. (2003a) except for a few minor features which appear to be
slightly different between the two reductions, for example, around
20.4~\AA\ and around 17.48~\AA\ in the rest frame of the source.
We attribute these discrepancies to the different binning methods
used and to the averaging of RGS1 and RGS2 in this work.  We also
note an unexplained $\sim 0.1$ \AA\ wavelength shift between the
Pounds et al. (2003a) data set and our reduction, but this does not
affect the analysis nor the main results.

\section{Data analysis}

\subsection{EPIC-pn spectrum}
\label{pnsec}

\begin{figure}
\centerline{\includegraphics[width=8.5cm]{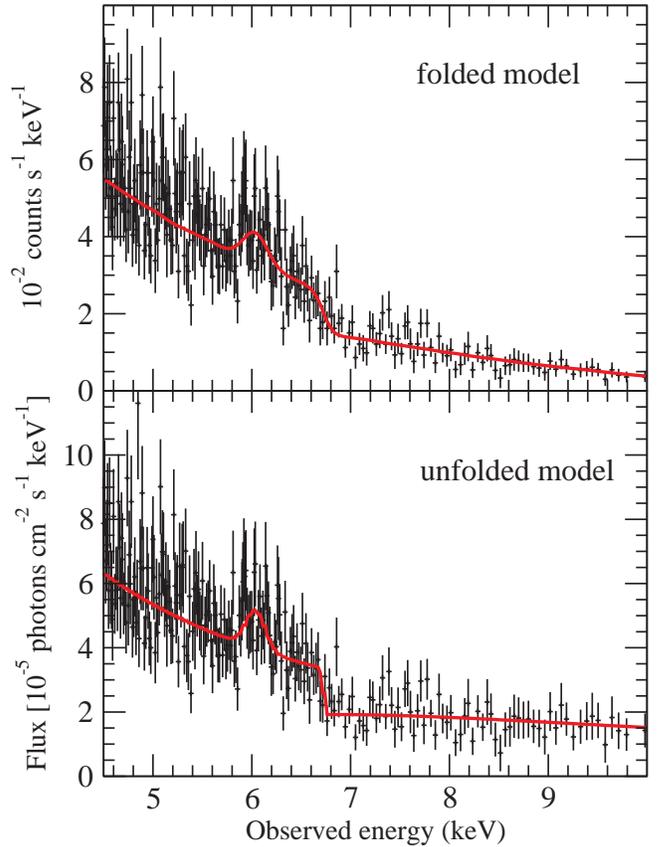}}
\caption{The EPIC-pn data and a simple fitted model, which is discussed
in \S~\ref{pnsec}. The upper panel shows the model folded through the
instrument response and compared with the data.  Bottom panel shows
the unfolded model.}
\label{pnmodel} 
\end{figure}

We use XSPEC (V.11.1.0) to fit the EPIC-pn data. We first fit the
(line-free) rest-frame 2--5 keV energy range with a simple power
law. The best fitted power law has a photon index of $\Gamma =
1.55\pm 0.05$ and a normalization of $(6.6\pm 0.4)\times10^{-4}$
ph\,cm$^{-2}$\,s$^{-1}$\,keV$^{-1}$ and gives $\chi^{2}_{\nu}=0.74$
for 487 degrees-of-freedom (d.o.f.). Extrapolating this power law
up to a rest-frame energy of 11~keV, we find a flux excess above the
power law at around 6.4~keV, which is indicative of an iron K$\alpha$
line, and a flux deficit below the power law at energies above 7~keV.
We add to the model a Gaussian emission line to account for the
Fe\,K$\alpha$ line and a photoelectric absorption edge to account
for the deficit. Fitting for all parameters simultaneously, we
find the best-fit Gaussian line center is at $6.04\pm 0.04$~keV (or
$6.53\pm 0.05$~keV in the rest frame) and a line width ($\sigma$) of
$0.096\pm 0.067$~keV. The total flux in the line is $(2.9\pm 1.4)\times
10^{-6}$ \phcms.  For the edge, we find a threshold energy of $6.72\pm
0.10$~keV, which is translated to a rest frame energy of $7.27\pm
0.11$~keV. The optical depth at the edge is $\tau=0.56\pm 0.10$. The
power law model with the Gaussian line and the absorption edge gives
a $\chi^{2}_{\nu}=0.983$ for 613 d.o.f. This model is plotted in
Figure~\ref{pnmodel} where we show the model, both folded through the
instrument and fluxed (i.e., unfolded).  We stress that this edge does
not necessarily contradict the presence of the line detected by Pounds
et al. (2003a), since K$\alpha$ edges have lines right next to them.

EPIC background spectra can occasionally have fluorescence
emission lines of Zn, Cu and Ni at energies above $\sim
7$ keV due to interaction of high energy particle with the
structure surrounding the detectors and the detectors themselves
(Lumb et al. 2002).\footnote{See also the \xmm\ User's Handbook: \\
http://xmm.vilspa.esa.es/external/xmm\_user\_support/documentation/uhb/node35.html}
These background emission-lines vary across the CCDs and are not
accounted for in the standard reduction process. We have verified that
this instrumental fluorescence radiation does not affect the source
or background regions we use. Nevertheless, on other regions of the
CCD these features are present and can lead to false identification of
narrow absorption lines in the source spectrum due to the background
subtraction.

\begin{figure*}
\centerline{\includegraphics[width=18cm]{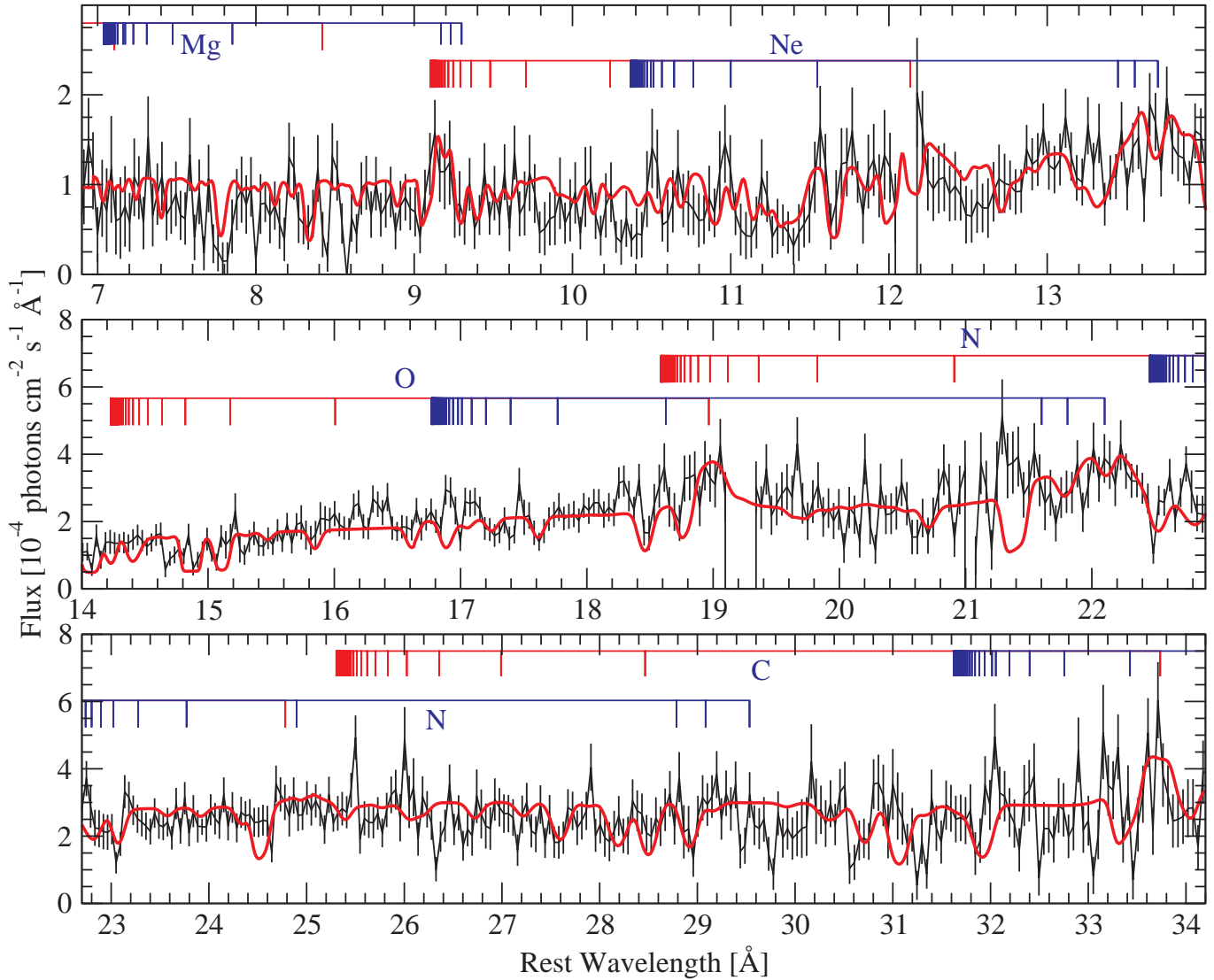}}
\vspace{-0cm}
\caption{Combined RGS1 and RGS2 spectrum of \pgt\ binned to $\sim 0.04$
\AA. The spectrum has been corrected for Galactic absorption and for
the redshift of the source. The rest-frame positions of lines from
H-like (red) and He-like (blue) ions of N, O, Ne, and Mg are marked
above the spectrum.  The lines from the lower ionization states of
O and Mg, and the L-shell lines of Fe are not marked.  Gaps in the
spectrum are due to chip gaps and have zero flux. The model discussed
in \S~\ref{rgsfit} is marked as the red curve.}
\label{pgtrgs} 
\end{figure*}

We also observe the lines at 2.68 keV and 1.47 keV claimed by Pounds et
al. (2003a) to be from S and Mg, only we identify them as different
lines at much lower velocities. The 2.68 keV line is identified
here as \ion{S}{15}\,He$\beta$ and the 1.47 keV line is identified
as \ion{Mg}{11}\,He$\beta$.

\subsection{RGS spectrum}
\label{rgsfit}

The combined RGS spectrum (RGS1 and RGS2) of \pgt\ is presented in
Figure~\ref{pgtrgs}. Numerous absorption lines and several emission
lines are detected. Table~\ref{abslines} lists the absorption lines
we identify with high certainty. For each line we list the wavelength 
at the source rest-frame ($\lambda_{\rm source}$), its equivalent
width (EW), the line identification, and the velocity blueshift. The
absorption lines were measured by fitting each line with a local
continuum and a Gaussian.  We also list the identification of the
lines by Pounds et al. (2003a) where available (see discussion below).
We identify K-shell lines of C, N, O and Mg and L-shell lines of O, Mg,
Si, Ar, and Fe.  The absorption line widths are consistent with the RGS
resolution, and with the present S/N we are not able to resolve the
intrinsic velocity widths.  In emission, we identify significantly
broadened lines of \ion{N}{7} Ly$\alpha$, \ion{O}{8} Ly$\alpha$,
the forbidden line of \ion{O}{7} and its He$\alpha$ resonance line,
the forbidden line of \ion{Ne}{9}, and the \ion{Mg}{11} He$\alpha$
resonance line, all in the rest frame of the source with no velocity
shift.

\begin{deluxetable*}{cclc|cclc}
\tablecolumns{8}
%\tabletypesize{\small}
\tabletypesize{\scriptsize}
%\rotate
\tablewidth{0pt}
\tablecaption{Identified absorption lines in the RGS spectrum
\label{abslines}}
\tablehead{
\multicolumn{4}{c|}{Present Work} &
\multicolumn{4}{c}{Pounds et al. (2003a)} \\
\colhead{$\lambda_{\rm source}$\tablenotemark{a} [\AA ]} &
\colhead{EW [m\AA ]} &
\colhead{Line ID [\AA ]} &
\multicolumn{1}{c|}{Velocity [\kms ]} & 
\colhead{$\lambda_{\rm source}$ [\AA ]} &
\colhead{EW [m\AA ]}  &
\colhead{Line ID [\AA ]} &
\colhead{Velocity [\kms ]}
 }
\startdata
$\phn 7.789\pm{0.033}$  & $127^{+35}_{-29}$  &  \ion{Mg}{11} He$\beta$ (7.851)   & $2400\pm{1300}$    & $\phn 7.80\pm0.15$ & $86\pm34$ & \ion{Mg}{12} Ly$\alpha$ (8.421) & $24300\pm 1000$ \\  
$\phn 8.316\pm{0.036}$  & $59^{+21}_{-16}$   &  \ion{Mg}{12} Ly$\alpha$ (8.421) &  $3700\pm{1300}$    &  \multicolumn{4}{c}{\nodata}               \\ 
$\phn 9.025\pm{0.034}$  & $41^{+30}_{-14}$ &\ion{Mg}{11} He$\alpha$ (9.169)\tablenotemark{b}&\nodata  & \multicolumn{4}{c}{\nodata} \\ 
$11.376\pm{0.037}$ & $185^{+57}_{-48}$  &  blend at $\sim 11.5$\tablenotemark{c} & $\sim 3230\pm{970}$& $11.17\pm0.03$ & $50\pm20$ & \ion{Ne}{10} Ly$\alpha$ (12.134) & $23700\pm 800$ \\ 
$11.649\pm{0.032}$ & $54^{+54}_{-16}$   &  blend at $\sim 11.8$\tablenotemark{d} & $\sim 3840\pm{810}$& \multicolumn{4}{c}{\nodata}  \\ 
\multicolumn{4}{c|}{\nodata}                                                                          & $12.40\pm0.05$ & $70\pm15$ & \ion{Ne}{9} He$\alpha$ (13.447) & $23400\pm1100$ \\
$14.902\pm{0.031}$ & $48^{+18}_{-11}$   &  \ion{Fe}{17} (15.014) &  $2240\pm{620}$       &              $14.78\pm0.07$ & $60\pm25$  & \ion{O}{8} Ly$\beta$ (16.006)  & $23000\pm 1300$ \\ 
$15.051\pm{0.032}$ & $32^{+13}_{-8}$    &  blend at $\sim 15.22$\tablenotemark{e} &  $\sim 3330\pm{630}$& \multicolumn{4}{c}{\nodata}  \\ 
$17.268\pm{0.040}$ & $46^{+25}_{-14}$   &  \ion{O}{7} He$\delta$ (17.396) &  $2210\pm{690}$           & $17.21\pm0.05$ & $25\pm10$  & \ion{O}{7} He$\beta$ (18.627) & $22900\pm 810$ \\ 
$17.621\pm{0.035}$ & $42^{+28}_{-12}$   &  \ion{O}{7} He$\gamma$ (17.768) &  $2480\pm{590}$           & $17.49\pm0.03$ & $120\pm25$  & \ion{O}{8} Ly$\alpha$ (18.969) & $23400\pm 470$ \\ 
$18.482\pm{0.032}$ & $61^{+23}_{-12}$   &  \ion{O}{7} He$\beta$ (18.627)  &  $2335\pm{520}$           &\multicolumn{4}{c}{\nodata}\\ 
$18.706\pm{0.035}$ & $47^{+23}_{-11}$   &  \ion{O}{8} Ly$\alpha$ (18.969) &  $4160\pm{550}$           &\multicolumn{4}{c}{\nodata}\\ 
\multicolumn{4}{c|}{\nodata}                                                                          & $19.94\pm0.05$ & $60\pm15$  & \ion{O}{7} He$\alpha$ (21.602) & $23100\pm 700$ \\
$22.488\pm{0.032}$&$40^{+17}_{-9}$   &  \ion{O}{4} K$\alpha$ (22.729, 22.777)&$3490\pm{420}$          &\multicolumn{4}{c}{\nodata}\\ 
$23.052\pm{0.032}$&$66^{+19}_{-15}$&\ion{O}{2} K$\alpha$ (23.302, 23.301, 23.3)&$3200\pm{410}$      & $22.93\pm0.03$ & $50\pm10$  & \ion{N}{7} Ly$\alpha$ (24.781)  & $22400\pm 360$ \\ 
$27.574\pm{0.037}$&$62^{+35}_{-15}$&  \ion{Ar}{11} (27.846, 27.881)&$3260\pm{400}$                    &\multicolumn{4}{c}{\nodata}\\ 
$28.528\pm{0.049}$ & $91^{+63}_{-27}$   &  \ion{N}{6} (28.780)   &  $2630\pm{510}$                    &\multicolumn{4}{c}{\nodata}\\ 
$28.948\pm{0.036}$ & $115^{+61}_{-27}$  &  \ion{Ar}{13} (29.209) &  $2680\pm{370}$                    &\multicolumn{4}{c}{\nodata}\\ 
$30.626\pm{0.037}$&$139^{+30}_{-25}$& \ion{Si}{12} (31.018, 31.027)\tablenotemark{f}&$\sim 3830\pm{360}$ &\multicolumn{4}{c}{\nodata}\\ 
$31.273\pm{0.036}$ & $162^{+62}_{-37}$  &  \ion{Ar}{12} (31.374)\tablenotemark{g}&    \nodata         & $31.10\pm0.03$ & $90\pm25$   & \ion{C}{6} Ly$\alpha$ (33.736) & $23300\pm 270$ \\ 
$33.461\pm{0.036}$ & $82^{+33}_{-24}$   & \ion{C}{6} Ly$\alpha$ (33.736)  &   $2450\pm{320}$          & \multicolumn{4}{c}{\nodata}      
\enddata
\tablenotetext{a}{Measured wavelength at the rest-frame of the source.}
\tablenotetext{b}{Blend with emission line which partially fills the trough.}
\tablenotetext{c}{Blend of \ion{Fe}{22} (11.427, 11.495), \ion{Fe}{23} (11.326, 11.319, 11.315, 11.423) and \ion{Ne}{9} (11.547) He$\beta$.}
\tablenotetext{d}{Blend of \ion{Fe}{22} (11.78) and \ion{Fe}{21} (11.825).}
\tablenotetext{e}{Blend of \ion{O}{8} Ly$\gamma$ (15.176) and \ion{Fe}{17} (15.261).}
\tablenotetext{f}{Blend with an unidentified line.}
\tablenotetext{g}{Not a certain identification.}
\end{deluxetable*}

In order to quantitatively explore the emission and absorption lines,
we have constructed a model for the entire RGS spectrum.  The present
method is an ion-by-ion fit to the data similar to the approach used
in Sako et al. (2001) and in Behar et~al. (2003). We first use the
continuum measured from the EPIC-pn data, but renormalized to the
RGS flux level. This continuum is then absorbed using the full set
of lines for each individual ion. Our absorption model includes
the first 10 resonance lines of H- and He- like ions of C, N, O,
Ne, and Mg as well as edges for these ions. The model also includes
our own calculation for the L-shell absorption lines of Fe (Behar et
al. 2001) as well as of Si, S, and Ar corrected according to laboratory
measurements (Lepson et al. 2003, 2005). Finally, we include inner-shell
K$\alpha$ absorption lines of O and Mg (Behar \& Netzer 2002), which
we detect in the spectrum. The absorbed spectrum is complemented by
the emission lines mentioned above, which are observed in the RGS
spectrum.\footnote{The atomic data and the code used for the model
are available upon request from the authors.}

By experimenting with the absorption line parameters, we find that
the observed lines are all blueshifted by about 3000~\kms\ with an
uncertainty of 500~\kms (see also Table~\ref{abslines}). In the model
we used a turbulence velocity of 1000~\kms\ to broaden the absorption
lines.  This width includes the instrumental broadening, which as
noted above, we could not separate from the intrinsic broadening.
Since the lines appear to be saturated, but no line goes to zero
intensity in the trough, we obtain the best fit by assuming a
covering factor of 0.7 for the X-ray continuum source. The best-fit
column densities that we find for the different ions are listed in
Table~\ref{coldens}. The emission lines are modeled using Gaussians
with uniform widths of $\sigma = 2500$ \kms\ (resolved, but again,
including the instrumental broadening), with no velocity shift,
and assumed to be unabsorbed. These lines at FWHM~$\simeq$ $6000\pm
1200$~\kms\ are even broader than those observed from the broad line
region in the visible band ($\sim 2000$ \kms ; Kaspi et al. 2000). The
model emission line fluxes are given in Table~\ref{emisflux}. The
entire best-fit spectrum is represented in Figure~\ref{pgtrgs} by the
red curve. The spectrum beyond 25~\AA\ is particularly challenging as
it comprises many unresolved lines from L-shell ions of Si, S, and Ar
while the RGS effective area drops rapidly. Several predicted lines
may be observed here (e.g., \ion{Ar}{13} - 28.92~\AA, \ion{Si}{12} -
30.71~\AA, \ion{Ar}{12} - 31.06~\AA, \ion{S}{13} - 31.93~\AA; these
wavelengths include the 3000 \kms\ shift). We are still unable to
explain several features seen in the data, e.g., around 8.5 \AA, 10.4
\AA, or 29.8 \AA , but the model gives a good fit to the data overall.

\begin{deluxetable}{lcc}
\tablecolumns{3}
%\tabletypesize{\small}
\tabletypesize{\scriptsize}
% \rotate
\tablewidth{150pt}
\tablecaption{Column densities for the RGS model
\label{coldens}}
\tablehead{
\colhead{Ion} &
\colhead{$N_{\rm ion}$ [\cmii ]} &
\colhead{Inferred $N_{\rm H}$ [\cmii ]} 
 }
\startdata
\ion{C}{6}    & $2\times10^{17}$  &  $1.0\times10^{21}$   \\    [0.2cm]
\ion{N}{7}   & $2\times10^{17}$  &  $2.2\times10^{21}$   \\
\ion{N}{6}    & $5\times10^{16}$  &  $5.6\times10^{20}$   \\     [0.2cm]
\ion{O}{8}  & $2\times10^{17}$  &  $5.1\times10^{20}$   \\
\ion{O}{7}   & $3\times10^{17}$  &  $7.7\times10^{20}$   \\
\ion{O}{6}    & $2\times10^{16}$  &  $1.4\times10^{20}$   \\
\ion{O}{5}     & $5\times10^{16}$  &  $3.4\times10^{20}$   \\
\ion{O}{4}    & $8\times10^{16}$  &  $5.4\times10^{20}$   \\
\ion{O}{3}   & $5\times10^{16}$  &  $3.4\times10^{20}$   \\
\ion{O}{2}    & $1\times10^{17}$  &  $6.8\times10^{20}$   \\     [0.2cm]
\ion{Ne}{10}    & $5\times10^{17}$  &  $5.0\times10^{21}$   \\
\ion{Ne}{9}   & $5\times10^{16}$  &  $5.1\times10^{20}$   \\    [0.2cm]
\ion{Mg}{12}  & $5\times10^{17}$  &  $1.6\times10^{22}$   \\
\ion{Mg}{11}   & $1\times10^{18}$  &  $3.3\times10^{22}$   \\
\ion{Mg}{10}    & $1\times10^{17}$  &  $8.9\times10^{21}$   \\
\ion{Mg}{9}   & $3\times10^{17}$  &  $2.6\times10^{22}$   \\
\ion{Mg}{8} & $1\times10^{17}$  &  $8.9\times10^{21}$   \\
\ion{Mg}{7}  & $1\times10^{17}$  &  $8.9\times10^{21}$   \\   [0.2cm]
\ion{Si}{12} & $2\times10^{17}$  &  $1.9\times10^{22}$   \\   [0.2cm]
\ion{S}{13} & $4\times10^{16}$  &  $8.2\times10^{21}$   \\   [0.2cm]
\ion{Ar}{13} & $2\times10^{16}$  &  $1.8\times10^{22}$   \\  
\ion{Ar}{12} & $3\times10^{16}$  &  $2.8\times10^{22}$   \\   
\ion{Ar}{11} & $5\times10^{16}$  &  $4.6\times10^{22}$   \\   [0.2cm]   
\ion{Fe}{17} & $3\times10^{17}$  &  $3.1\times10^{22}$   \\
\ion{Fe}{8}& $2\times10^{17}$  &  $2.1\times10^{22}$   \\
\ion{Fe}{19}  & $3\times10^{16}$  &  $3.1\times10^{21}$   \\
\ion{Fe}{20}   & $5\times10^{16}$  &  $5.2\times10^{21}$   \\
\ion{Fe}{21}  & $1\times10^{17}$  &  $1.0\times10^{22}$   \\
\ion{Fe}{22} & $3\times10^{17}$  &  $3.1\times10^{22}$   \\
\ion{Fe}{23}& $1\times10^{17}$  &  $1.0\times10^{22}$   \\
\ion{Fe}{24} & $1\times10^{17}$  &  $1.0\times10^{22}$   
\enddata
\end{deluxetable}
\section{Interpretation and Discussion}

Pounds et al. (2003a) detected eight absorption lines in the same RGS
spectrum (see their Table~2). We list their detections (including
one line from their Table~1) in Table~\ref{abslines}. The EW and
wavelengths given by Pounds et al. (2003a) are presumably from
their globally fitted model, thus there are certain discrepancies
between them and our {\it measured} quantities. We note again the
small wavelength shift of about 0.1~\AA\ between the data shown here
and the Pounds et al. (2003a) data. This does not affect our main
results  as it can add only about 1500 \kms\ (at $\sim 20$~\AA )
to the outflow velocity we find, which cannot explain the order of
magnitude difference in the outflow velocity.  We note that the
velocities obtained in Pounds et al. (2003a) from the EPIC data
(their Table~1) are in err. These results have been revised in the
erratum by Pounds et al. (2005).\footnote{See another revised version
of Pounds et al. (2003a) at http://arXiv.org/abs/astro-ph/0303603
.} Even in Pounds et al. (2005) the velocities span a wide range
from 22600~\kms\ to 32700 \kms , which is weakly consistent with the
24000 \kms\ velocity claimed from the RGS data (Pounds et al. 2003a).
Attempting to model the data with a 24000\,\kms\ model, we were not
able to reproduce all the absorption lines in Table~\ref{abslines}.
Although it may be possible that both velocities are present,
we suspect that most of the column density lies in the 3000 \kms\
component, which we discuss below.

An outflow velocity of 3000 \kms\ places \pgt\ at the high-velocity
end of, or even slightly faster than, typical Seyfert~1 outflows. The
more interesting parameters of the outflow such as its location and
the associated mass outflow rate are much harder to obtain by means of
mere absorption spectroscopy. For an extended ionization cone geometry,
the distance of the base of the cone from the continuum source for
a given ionization species formed at an ionization parameter $\xi =
L/(n_e r^2)$ can be roughly estimated by (e.g., Behar et al. 2003):
\begin{equation}
r_{min} \la \frac{n_H}{n_e}\frac{L}{\xi}\frac{1}{N_{\rm H}}= \hspace{5.9cm}
\end{equation}
\begin{equation*}
\hspace{0.4cm} = 27 \left( \frac{L}{10^{44} {\rm\,erg\,s}^{-1}} \right)
\left( \frac{100 {\rm\,erg\,s^{-1}\,cm}}{\xi} \right)
\left( \frac{10^{22} {\rm\,cm}^{-2}} {N_{\rm H}} \right)  {\rm pc.}
\end{equation*}
The standard hydrogen to electron density ratio $n_H/n_e$ in a
fully ionized astrophysical plasma is 0.83. With an ionizing
luminosity, $L$, of a few times 10$^{44}$ erg~s$^{-1}$, this places
the typical ionization component of the \pgt\ absorber (e.g.,
$\xi\sim 50$\,erg\,s$^{-1}$\,cm for \ion{O}{8}) on large scales,
much further out than the estimates of Pounds et al. (2003a) who
invoke an opaque photosphere (King \& Pounds 2003)
on scales of $10^{15}$\,cm.

\begin{deluxetable}{lc}
\tablecolumns{2}
% \tabletypesize{\footnotesize}
% \tabletypesize{\scriptsize}
% \rotate
\tablewidth{150pt}
\tablecaption{Model Broad\tablenotemark{a} Emission Line Fluxes
\label{emisflux}}
\tablehead{
\colhead{Line [\AA ]} &
\colhead{Flux [$10^{-5}$ \phcms]} 
}
\startdata
\ion{C}{6}   Ly$\alpha$ (33.737)    & 8.8 \\
\ion{N}{7}  Ly$\alpha$ (24.782)    & 4.2 \\
\ion{O}{7}  forbidden  (22.101)    & 8.3 \\
\ion{O}{7}  He$\alpha$ (21.602)    & 3.2 \\
\ion{O}{8} Ly$\alpha$ (18.969)    & 5.9 \\
\ion{Ne}{9}  forbidden  (13.699)    & 2.0 \\
\ion{Ne}{10}   Ly$\alpha$ (12.134)    & 1.3 \\
\ion{Mg}{11}  He$\alpha$ (9.169)     & 1.3 
\enddata
\tablenotetext{a}{FWHM = 6000\,\kms .}
\end{deluxetable}

The total mass outflow rate of the wind is impossible to estimate
without knowing the opening angle, $\Omega$ (sky covering fraction), of
the wind. One way to obtain $\Omega$ is to measure the volume emission
measure of narrow emission lines assumed to emerge from the absorbing
outflow (e.g., Behar et al. 2003). In the present spectrum, however,
we do not detect any narrow emission lines, but only broad lines,
which probably originate in the inner BLR.  The apparent absence of
narrow emission lines could imply that the opening angle is rather
small and perhaps so is the mass outflow rate. Pounds et al. (2003a)
obtain much higher column densities $\sim$~10$^{24}$~cm$^{-2}$ than we do
and their spherically symmetric expanding photosphere results in a high
estimate of the mass outflow rate: $3M_\sun$ yr$^{-1}$.

McKernan et al. (2004) suggest that some portion of the absorption
detected by Pounds et al. (2003a) could be due to hot local
intergalactic medium (IGM) since the 24000 \kms\ outflow velocity
coincides with the recession velocity of \pgt . Our interpretation,
of a 3000 \kms\ outflow, is not consistent with the IGM interpretation.
Also, the high column density inferred, as well as the high ionization
level observed, are not typical of the IGM. Finally no local ionized
absorber is detected in the UV towards this source.

Although the present model does not produce a perfect fit to
the observed spectrum, it does explain self consistently the
absorption lines with a single outflow velocity of 3000 \kms . We
stress that our method is not a global fit and that all of the ions
appearing in Table~\ref{coldens} have been identified directly in the
spectrum. The inclusion of emission lines in our model explains why
some Ly$\beta$-like and He$\beta$-like lines appear to be deeper than
their corresponding (saturated) Ly$\alpha$ and He$\alpha$ lines. This
happens due to filling of the absorption trough by the emission. For
example, the \ion{Mg}{11}~He$\alpha$ absorption line is filled by
emission which causes the \ion{Mg}{11} He$\beta$ absorption line to
appear deeper (see Figure~\ref{pgtrgs}).

As specified in Table~\ref{coldens}, the best-fit ionic column
densities are between 2$\times$10$^{16}$~\cmii\ and 10$^{18}$~\cmii .
For O and Mg, for which both K-shell and L-shell ions are present,
the L-shell ions are somewhat less abundant in the X-ray absorbing
plasma, as expected. We estimate the measured column densities to be
accurate to within a factor of a few. Assuming that the absorption by
each ionic species is produced where it attains its maximal fractional
abundance, we estimate the effective hydrogen column density, $N_{\rm
H}$, by ascribing crude, peak ionic fractional abundances of 0.8
and 0.3 to K-shell and L-shell ions, respectively (e.g., Kallman et
al. 2004), and by assuming solar abundances of (2.45, 1.12, 4.90,
1.23, 0.380, 0.355, 0.162, 0.0363 and 0.320)$\times 10^{-4}$ for C,
N, O, Ne, Mg, Si, S, Ar, and Fe, respectively.  This yields inferred
$N_{\rm H}$ values for each element that are consistent to an order of
magnitude. Discrepancies between elements may reflect deviations from
solar abundances.  We find that the ionic column densities in our model
correspond to a hydrogen column of roughly $10^{21}$--10$^{22}$~\cmii\
(Table~\ref{coldens}), similar to typical values measured for Seyfert
galaxies.  This hydrogen column density is two orders of magnitude
smaller than the Pounds et al. (2003a) result.

We believe the detected span of charge states is primary a density
effect rather than a radial stratification effect, due to the uniform
outflow velocity and the similar inferred hydrogen column density,
$N_{\rm H}$, for all the ions.  The observed ionic species reflect a
broad and rather flat ionization distribution along the line of sight,
again, an effect which has been observed in many Seyfert~1 outflows
(e.g., Kaspi et~al. 2001) as well as in Seyfert~2 galaxies (Kinkhabwala
et~al. 2002). In the latter case, all charge states appeared to be
present at similar distances from the central source, which implies
local density gradients of 3--4 orders of magnitude. In this sense and
to the end that our diagnostics of a 3000~\kms\ outflow is correct,
\pgt\ resembles a Seyfert~1 galaxy, only the wind is slightly faster,
perhaps a consequence of its higher luminosity.

Tumlinson et al. (2005) study the UV spectrum of \pgt\ and find two
absorption systems blueshifted by $\sim 5000$ and $\sim 9000$ \kms\
relative to the rest frame. They attribute them to IGM in connection
with two galaxy groups on the line of sight to \pgt . Although
the absorption is identified also in the \ion{O}{6} UV lines the column
densities they measure are too low for an X-ray detection of these IGM
components. We examined the {\it Far Ultra Violet Explorer (FUSE)}
observation of \pgt\ from the {\it FUSE} archive taken 14 months
before the \xmm\ observation. We find no evidence for absorption
in \ion{O}{6} at the outflow velocities of $\sim 24000$ \kms\ or
$\sim 3000$ \kms .  With the ionic column density from our RGS model
such lines should have been detected. The absence of the \ion{O}{6}
absorption in the {\it FUSE} spectrum could be because of changes in
the absorber between the two observations. An alternative explanation
is that the strong K$\alpha$ \ion{O}{6} line at 21.79 \AA\ required
by our model is confused with the Galactic K$\alpha$ \ion{O}{1}
absorption line which is expected at 21.76 \AA\ in the rest-frame
wavelength scale of Figure~\ref{pgtrgs}.  The \ion{O}{6} emission
lines in the {\it FUSE} spectrum appear to have a narrower component
with FWHM of about 2000 \kms , which correspond to the width of the
Balmer and the \ion{C}{4} emission lines (e.g., Baskin \& Laor 2005)
but also a broader component consistent with the FWHM of 6000~\kms\
we find for the X-ray emission lines.

The observed Fe-K absorption edge can be understood in the context
of the ionized outflow. Absorption spectra of Fe ions feature edges
due to ejection of K-shell electrons. For \ion{Fe}{1} to \ion{Fe}{24}
these edges cover the range from 7.1 to 9 keV. In the more ionized
cases, the spectral structure is not a simple edge as demonstrated
most recently by Kallman et~al. (2004). However, for ions close to
neutral, an edge is an excellent approximation.  The edge energy we
find in the EPIC-pn spectrum (Figure~\ref{pnmodel}) falls at $7.27\pm
0.11$~keV in the rest frame of \pgt. This corresponds to charge states
of \ion{Fe}{4} to \ion{Fe}{10}.  If the Fe edge is also due to the same
3000~\kms\ outflow observed with the RGS then it is associated with
even less ionized gas. The K-edge position of the low charge states
of Fe all fall close to each other and certainly can not be resolved
with the current CCD instruments. In addition, the broad ionization
distribution in the outflow suggests that the observed edge is due
to several charge states. Indeed, the high optical depth at the edge
(0.56) requires the contribution of several (5--10) Fe ions, with
a column density of about 10$^{18}$~\cmii\ each. The more highly
ionized Fe ions (Table~\ref{coldens}) also must contribute to the
photoelectric absorption at higher energies. We note that even deeper
edges have been measured for similar sources (e.g., Boller et~al. 2003).

\section{Conclusions}

\begin{enumerate}
\renewcommand{\theenumi}{(\arabic{enumi})}

\item We have provided a self consistent model to the ionized outflow
of \pgt\ revealing an outflow velocity of approximately 3000~\kms.
Our model reproduces many absorption lines in the RGS band, although
the S/N of the present data set is rather poor and some of the noise
might be confused with absorption lines.

\item The present approach is distinct from the commonly used global
fitting methods and also from the line-by-line approach used by
Pounds et al. (2003a). It allows for a physically consistent
fit to the spectrum and is particularly appropriate for a
broad-ionization-distribution absorber as observed here for \pgt .

\item The present model also features several broad (FWHM = 6000~\kms)
emission lines, which are observed directly in the data.

\item A broad and relatively flat ionization distribution is found
throughout the X-ray outflow consistent with a hydrogen column density
of roughly $10^{21}$--$10^{22}$ \cmii. This is reminiscent of the outflow
parameters measured in other well studied Seyfert galaxies.

\item We also detect Fe-K absorption, which was identified by Pounds
et~al. (2003a) as a strongly blueshifted \ion{Fe}{26} absorption
line. We find that most of the Fe-K opacity can alternatively be
attributed to several consecutive, low charge states of Fe, although
it can not be assessed whether the absorber is co-moving with the
outflow or not. Future missions with microcalorimeter spectrometers
on board might be able to address this interesting question.

\end{enumerate}

\acknowledgments

We thank D. Chelouche and A. Laor for helpful discussions.  We are
grateful for several valuable suggestions by an anonymous referee.
This research was supported by the Israel Science Foundation (grant no.
28/03), and by a Zeff fellowship to S.K.

\end{document}